\documentclass[conference]{IEEEtran}
\usepackage{cite}
\usepackage{amsmath,amssymb,amsfonts}
\usepackage{algorithmic}
\usepackage{graphicx}
\usepackage{textcomp}
\usepackage{xcolor}
\usepackage[most]{tcolorbox}
\usepackage{fancyhdr}
\usepackage[ngerman,english]{babel}
\usepackage[T1]{fontenc}
\usepackage[utf8]{inputenc}
\usepackage{listings}
\lstdefinelanguage{diff}{
sensitive=true,
morecomment=[f][\color{gray}][0]{diff},
morecomment=[f][\color{gray}][0]{index},
morecomment=[f][\color{blue}][0]{@@},
morecomment=[f][\color{magenta}][0]{***},
morecomment=[f][\color{violet}][0]{!},
morecomment=[f][\color{red}][0]-,
morecomment=[f][\color{green!60!black}][0]+,
morecomment=[f][\color{magenta}][0]{---},
morecomment=[f][\color{magenta}][0]{+++},
morecomment=[f][\color{gray}][0]{Binary},
morecomment=[f][\color{gray}][0]{Only},
morecomment=[f][\color{gray}][0]{old},
morecomment=[f][\color{gray}][0]{new},
morecomment=[f][\color{gray}][0]{rename},
morecomment=[f][\color{gray}][0]{similarity},
morecomment=[f][\color{gray}][0]{deleted},
morecomment=[f][\color{magenta}][0]{***************},
morecomment=[f][\color{red}][0]<,
morecomment=[f][\color{green!60!black}][0]>,
morecomment=[f][\color{blue}][0]{0},
morecomment=[f][\color{blue}][0]{1},
morecomment=[f][\color{blue}][0]{2},
morecomment=[f][\color{blue}][0]{3},
morecomment=[f][\color{blue}][0]{4},
morecomment=[f][\color{blue}][0]{5},
morecomment=[f][\color{blue}][0]{6},
morecomment=[f][\color{blue}][0]{7},
morecomment=[f][\color{blue}][0]{8},
morecomment=[f][\color{blue}][0]{9},
}[comments]
\usepackage{scrhack}
\usepackage{svg}
\usepackage{svg-extract}
\usepackage{float}
\usepackage{rotating,longtable}
\usepackage[skipabove=10pt,skipbelow=10pt]{mdframed}
\usepackage{fancyvrb}
\newenvironment{borderedframe}[1]
{\mdfsetup{
frametitle={\colorbox{white}{\space#1\space}},
innertopmargin=0pt,
frametitleaboveskip=-0.8\ht\strutbox,
frametitlealignment=\center,
skipabove=0pt,
skipbelow=0pt
}
\begin{mdframed}%
}
{\end{mdframed}}

\usepackage{tabularx}
\usepackage{booktabs,graphicx}
\usepackage{float}
\usepackage{enumitem}
\usepackage{array}
\usepackage{tablefootnote}
\usepackage{paralist}
\usepackage{nicematrix,enumitem,booktabs}
\usepackage{array}
\newcolumntype{L}[1]{>{\raggedright\let\newline\\\arraybackslash\hspace{0pt}}m{#1}}
\newcolumntype{C}[1]{>{\centering\let\newline\\\arraybackslash\hspace{0pt}}m{#1}}
\newcolumntype{R}[1]{>{\raggedleft\let\newline\\\arraybackslash\hspace{0pt}}m{#1}}

\PassOptionsToPackage{hyphens}{url}
\usepackage{hyperref}

\newcommand{\kuerzungskandidat}[1]{#1}
\newtcolorbox{observation}{colback=gray!5!white, boxrule=0.1mm, arc=1mm, boxsep=1mm, left=1mm, right=1mm, top=1mm, bottom=1mm}
\begin{document}

\title{A Full-fledged Commit Message Quality Checker Based on Machine Learning}

\author{\IEEEauthorblockN{David Faragó}
\IEEEauthorblockA{\textit{Innoopract GmbH \& QPR Technologies} \\
Karlsruhe, Germany \\
farago@qpr-technologies.de}
\and
\IEEEauthorblockN{Michael Färber}
\IEEEauthorblockA{\textit{Karlsruhe Institute of Technology} \\
Karlsruhe, Germany \\
michael.faerber@kit.edu}
\and
\IEEEauthorblockN{Christian Petrov}
\IEEEauthorblockA{\textit{Innoopract GmbH} \\
Karlsruhe, Germany \\
cpetrov@innoopract.com}
}

\maketitle
\thispagestyle{fancy}

\begin{abstract}

Commit messages (CMs) are an essential part of version control.
By providing important context in regard to what has changed and why, they strongly support software maintenance and evolution.
But writing good CMs is difficult and often neglected by developers. 
So far, there is no tool suitable for practice that automatically assesses how well a CM is written, including its meaning and context.
Since this task is challenging, we ask the research question: how well can the CM quality, including semantics and context, be measured with machine learning methods?
By considering all rules from the most popular CM quality guideline, creating datasets for those rules, and training and evaluating state-of-the-art machine learning models to check those rules,
we can answer the research question with: sufficiently well for practice, with the lowest F$_1$ score of 82.9\%, for the most challenging task.
We develop a full-fledged open-source framework that checks all these CM quality rules.
It is useful for research, e.g., automatic CM generation, but most importantly for software practitioners to raise the quality of CMs and thus the maintainability and evolution speed of their software.

\end{abstract}


\section{Introduction}
\label{sec:introduction}

\textbf{Motivation.} Although code should best be self-explanatory, it is unable to contain all context,
such as the implementation decisions (e.g., technical trade-offs) or the reasons for the code change (e.g., the business requirement or bug report motivating the commit).
We define context as all relevant information that the code change does not convey by itself.
CMs of code repositories that document this context are the most important way to understand the code change~\cite{Huang2020learning} and therefore help future development, evolution, and maintenance.

\begin{table}[h]
  \caption{\label{table:chrisBeamsRules}Chris Beams' CM quality guideline \cite{Beams}: rules and examples}
 \centering
\setlength{\tabcolsep}{8pt}
\begin{tabular}{|lll|}
\toprule
\textbf{Rule} & \textbf{Description} & \textbf{L}\hyperlink{level}{$^1$}\\
\midrule
R1&Separate subject from body with a blank line&F\\
R2&Limit the subject line to 50 characters&F\\
R3&Capitalize the subject line&F\\
R4&Do not end the subject line with a period&F\\
R5&Use the imperative mood in the subject line&SY\\
R6&Wrap the body at 72 characters&F\\
R7&Use the body to explain what and why vs. how&SE\\
\bottomrule
\end{tabular}
\begin{borderedframe}{R5 violated (``fix'' used as a noun, not as verb)}
\begin{Verbatim}[fontsize=\scriptsize]
Linter error fix
\end{Verbatim}
\end{borderedframe}
\begin{borderedframe}{R7 violated (describes "how", but not "what" and "why")}
\begin{Verbatim}[fontsize=\scriptsize]
Duplicate zval before add_next_index_zval
\end{Verbatim}
\end{borderedframe}
\begin{borderedframe}{R7 violated (unclear what was wrong before the change\textsuperscript{a})}
\begin{Verbatim}[fontsize=\scriptsize]
Fix Sass + CSS Modules (#3186)
\end{Verbatim}
\end{borderedframe}
\begin{borderedframe}{R5, R7 satisfied (simple change, context sufficient for R7)}
\begin{Verbatim}[fontsize=\scriptsize]
Fix linter errors
\end{Verbatim}
\end{borderedframe}
\begin{borderedframe}{R5, R7 satisfied}
\begin{Verbatim}[fontsize=\scriptsize]
Fix running ALTER TABLE statements in Execute SQL tab

This fixes a bug introduced in 73efa11. Because SQLite
reports ALTER TABLE statements to return one column
worth of data, DB4S assumed they are close to a SELECT
statement and therefore did not fully execute them.

See issues #2563 and #2622.
\end{Verbatim}
\end{borderedframe}
\textsuperscript{a}``No module-specific actions (like compiling the classnames) got done.''\cite{snowpack3186}\\[-1ex]
\end{table}
\addtocounter{footnote}{1}
\footnotetext{\label{orga3451ac}\hypertarget{level}Abstraction level F: formatting, SY: syntax, SE: semantics}

A comprehensive yet easy to understand CM history is one of the most powerful maintenance approaches~\cite{Beams} and
strongly benefits the code comprehension and team communication,
especially with today's increase in remote work~\cite{Motta2018}.
Despite these advantages, CMs are %
cultivated by only few software developers  (see Sec.~\ref{sec:discussion} and~\cite{Jiang2017a,nie2021coregen}) --
or in Linus Torvalds' words: \href{https://github.com/torvalds/linux/pull/17/\#issuecomment-5659933}{``GitHub is a total ghetto of crap commit messages''}.
This is confirmed by our experiment with 5,000 random CMs from GitHub:
our framework (see Sec.~\ref{sec:approach}) assesses 90\% of the CMs as low quality, and 55\% as low quality due to missing context.
In contrast, for the well-maintained~\cite{LinuxConvention} CM history of the Linux Kernel, our framework assesses only 3\% of the CMs as low quality due to missing context. 
Applied broadly, e.g., as automatic CM quality reviewing tool run locally or as part of a CI pipeline in the cloud, a CM quality checker could train developers and lead to CM histories
with much higher quality, 
resulting in more maintainable and faster evolving software. 

\textbf{CM quality guideline.}
Developers that cultivate CMs with high quality often follow certain rules.
Table~\ref{table:chrisBeamsRules} gives a summary and examples for \cite{Beams},
the most prominent (see Sec.~\ref{sec:relatedwork}) CM quality guideline.
It sums up most ideas from 
older guidelines
into a 7-rule convention, establishing etiquette for formatting, syntax, semantics, and contextual information:
The formatting rules R1 to R4, R6 help readability of CMs and are trivial to verify in an automated way.
The remaining two rules require capable natural language understanding (e.g., for the examples in Table~\ref{table:chrisBeamsRules}):
Syntactic rule R5 enforces imperative verb mood, which improves consistency since it is used in CMs generated by git, too.
R5 also improves understandability because the mood distinguishes descriptions about what the change does
from descriptions about the context and the way things worked before the change.
E.g., for the CM ``Incorrect changes were stored'', it is not clear whether storing incorrect changes
is the fix or the undesired behavior before the fix.
Semantic rule R7 says that a CM should leave out details about ``how`` a change has been made since code should be self-explanatory,
and rather focus on ``what`` has changed (i.e. summary for understanding) and ``why`` the change has been made in the first place
(reasons for the code change),
which is particularly useful for software maintainability and evolution.
Hence, R7 demands that the CM provides the relevant context, and thus requires investigating the CM's meaning.
The amount of required context depends on the situation:
Beams argues that ``a single line is fine, especially when the change is so simple that no further context is necessary''.
One-line CMs are very common, making up 35 \% of our samples.

\textbf{Contributions.} 
So far, there is no practical approach to check CMs on a semantic level (see Sec.~\ref{sec:relatedwork}).
This article fills this gap and presents a practical and programming language-agnostic framework (see Sec.~\ref{sec:approach}) integrating state-of-the-art NLP methods to rigorously check all of Beams' guideline.
We determined Beams' guideline as gold standard by surveying research articles and CM guidelines (see Sec.~\ref{sec:relatedwork}).

We (two authors and further three experienced software developers) create four labeled datasets (see Sec.~\ref{sec:orgd1c817e})
to train and evaluate state-of-the-art machine learning models for our framework (see Sec.~\ref{sec:org68dea4d}):
our models achieve state-of-the-art performance in assessing the CM quality on the levels format,
syntax (F$_1$ score of 97.8\%), and semantics (F$_1$ score of 82.9\%).

So our contributions are datasets to train machine learning (ML) models, evaluations of ML models, and a framework based on ML to automatically
check CM quality according to Beams' guideline. We also implemented a tool based on our framework, enabling everyone to assess commit message quality, both locally and using GitHub workflows. Our contributions are open-sourced\footnote{\label{foot:material}\hypertarget{sp}See \url{https://doi.org/10.6084/m9.figshare.22096736} and \url{https://github.com/commit-message-collective/beams-commit-message-checker}}.

\section{Related Work}
\label{sec:relatedwork}

\textbf{CM quality assessment.}
Table~\ref{fig:table:commitMessageQualityAssessment} summarizes related work on assessing CM quality.
It contains a lot of gray literature:  %
software guides, style guides, blog articles, wiki pages,
guidelines with criteria by which to assess the CM quality
-- there is no formal standard on CM quality.

We pick the guideline~\cite{Beams} by software practitioner Chris Beams for our work because of three reasons:
(1) It is the most popular: It is cited in about 10\% of the repositories having contribution rules (a \verb!CONTRIBUTING.md! file).
Some of the contribution rules do not cite \cite{Beams}, but require good CM semantics more vaguely,
e.g., that the CM be ``understandable'', ``meaningful'', or ``well documented''.
None of the repositories with contribution rules contradict the semantic rule R7.
The only other cited CM quality guideline is \cite{Pope},
in about 6\% of those repositories.
(2) It is the most complete: it covers all CM aspects mentioned by other guidelines, besides minor variations in formatting rules.
(3) It is cited by all five research articles about CM quality assessment.

\begin{observation}
\textbf{Insight 1:} Chris Beams' article is the most comprehensive guideline on commit message quality. In our study, we evaluate commit message quality based on its rules.
\end{observation}

\begin{table}[tb]
\caption{\label{fig:table:commitMessageQualityAssessment}CM quality assessment: related work}
\centering
\begin{small}
\begin{tabular}{rlllll}
\toprule
\textbf{Year} & \textbf{Source} & \textbf{Type} & \textbf{Level}\hyperlink{level}{$^1$} & \textbf{AE}\tablefootnote{\label{orgc80e499}Automated evaluation: evaluates criteria in an automated way.}\\
\midrule
2022 & \cite{Tian22} & Conference article & F, SY, SE & \checkmark \\
2021 & \cite{Itk2021} & Software guide & F & \checkmark \\
2020 & \cite{Dioptra2020} & Style guide & F, SY, SE & \\
2020 & \cite{Schweizer2020} & Master thesis & F, SY, SE & \\
2019 & \cite{Alberto} & Master thesis & F, SY & \checkmark\\
2019 & \cite{BadCommitMessageBlocker} & Tool website & F, SY & \checkmark\\
2019 & \cite{Timescale} & Tool website & F, SY & \checkmark\\
2018 & \cite{Chahal2018} & Journal article & F, SY, SE & \\
2017 & \cite{LinuxConvention} & Software guide & F, SY, SE & \\
2017 & \cite{Conventional} & Style guide & SY, SE & \\
2016 & \cite{GitCommitBear} & Tool website & F, SY & \checkmark\\
2016 & \cite{Gitmoji} & Style guide & F, SE & \\
2015 & \cite{agrawal2015commit} & Workshop article & - &\\
2014 & \cite{Beams} & Blog article & F, SY, SE & \\
2014 & \cite{GitScm} & Book & F, SY, SE & \\
2013 & \cite{Hearth13} & Blog article & F, SE & \\
2013 & \cite{SOCommit} & Wiki page & F, SY, SE & \\
2012 & \cite{ErlangOtpWritingGoodCommitMessages} & Wiki page & F, SY, SE & \\
2011 & \cite{firstUsagesCommitConventionAngular} & Software guide & F, SY, SE & \\
2009 & \cite{WhoT} & Blog article & F, SY, SE & \\
2008 & \cite{Pope} & Blog article & F, SY, SE & \\
2000 & \cite{Mockus2000a} & Journal article & - & \\
\bottomrule
\end{tabular}
\end{small}
\end{table}

There is only little research (five articles) in the related work. 
Only the recently published \hypertarget{tian-et-al}{Tian et al.}~\cite{Tian22} is closely related to our work:
They analyze the reason for a CM to be ``good'',
develop a corresponding taxonomy about how CMs convey ``what'' and ``why'' information,
and train machine learning models to automatically classify CMs as good.
Their best model has a high performance: 77.6\% positive class F$_1$, 73.9\% negative class F$_1$.
However, they focus on a thorough theoretic analysis about the taxonomy of commit messages instead of
realizing a checker for a CM's usefulness in practice, e.g. according to a guideline like \cite{Beams}.
They (2 experts) only investigate a low number (5) of projects with little variance: all are high quality open-source Java projects about web communication.
Furthermore, for a large portion of CMs, they rely on unsuitable pattern matching:
\begin{compactitem}
\item Any CM with automatically generated parts is excluded,
  e.g., CMs containing \verb!pull request #! (15\% to 30\% of all CMs).
  This is unsuitable since CMs containing automatically generated parts can also be of
  low quality and should thus be improved, while others do contain (automatically generated or manually added)
  text making them high quality CMs.
\item The ``why'' criterion is already fulfilled if the CM contains an
  issue- or PR-link (about 40\% of all CMs), or the word ``fix'' (about 30\% of all CMs).
  Relevant information hidden behind issue- or PR-links is unsuitable, as described in Sec.~\ref{sec:threats}.
  Pattern matching the word ``fix'' is unsuitable as this is insufficient to check what R7 demands:
  that the CM makes clear the reasons why the change was made,
  clarifying the way things worked before the change and what was wrong with that (see Table \ref{table:chrisBeamsRules}).
\end{compactitem}
In contrast, we (5 experts) investigate 427 projects with high variance
(multiple domains, programming languages, cultures), also consider CMs with
automatically generated parts, ignore issue- and PR-links, and semantically
check whether a CM really conveys what and why vs how.

\begin{observation}
  \textbf{Insight 2:} The only study on the most challenging task of assessing semantic commit message quality (R7 of Beams' guideline) falls short in fully capturing semantics and uses a dataset of limited diversity. We account for those shortcomings.
\end{observation}
The other four research articles are still related, but less:
Chahal et al. \cite{Chahal2018} conduct a multi-vocal literature review including gray literature.
They determine 11 criteria to measure CM quality and rank the importance of the chosen metrics through an expert survey.
These metrics are well covered by \cite{Beams}: the three most important metrics are variations of R5 and R7, and the atomicity rule
that a commit should have one logical change. But atomicity is an attribute of the commit, not its message.
Schweizer~\cite{Schweizer2020} investigates the fluctuations in quality metrics in commit histories.
They use~\cite{Beams} as metric for the CM quality.
Mockus et al. \cite{Mockus2000a} analyze commits from a quality perspective by considering commit metadata
like number of unique CMs and size of commit comments (smaller messages indicating immaturity).
Agrawal et al. \cite{agrawal2015commit} study 
literature to operationalize CM quality metrics.
Their Google search for “good commit logs” yields more than 57 million search results.
They use the top five hits, which are all gray literature: besides \cite{Hearth13,Pope,ErlangOtpWritingGoodCommitMessages,SOCommit}, \cite{Beams} is one of them.

\textbf{Automatic CM quality assessment.}
We found six sources that deal with the automated evaluation of CM quality:
The only research paper is \cite{Tian22}, which contains a model for automation, but is of limited practical use (see above).
Alberto et al. \cite{Alberto} propose a tool to validate a syntactic interpretation of Chris Beams' rules.
The other four sources are tools:
The open-source tool GitCommitBear ~\cite{GitCommitBear} automatically analyzes R5 by a syntactic analysis
based on classical natural language processing, which faces performance issues.
The other three tools try to cover all of \cite{Beams} but fail on R7:
The open-source medical image processing (ITK) software guide~\cite{Itk2021}
checks variations of Beams' formatting rule R1, R2, R6.
The open-source project~\cite{Timescale} %
validates all syntactic rules of Chris Beams and heuristically checks R5 using a keyword-based approach
(for R7, the script performs no check and only states ``Not enforceable'').
The open-source project~\cite{BadCommitMessageBlocker} validates all of Chris Beams' rules but R7.
For R5 they employ a similar technique to \cite{GitCommitBear} and face similar issues.
They state that R7 is too subjective, but we show a substantial agreement between experts assessing R7 in Section~\ref{sec:orgd1c817e}.

\textbf{Automatic generation of CMs.}
Most research about CMs is dealing with their automatic generation out of code changes,
without considering the intent behind the code changes~\cite{Huang2020learning}:
An automatically generated CM is evaluated using Natural Language Generation metrics such as BLEU,
which measure its closeness to CMs that humans generated out of the code changes~\cite{papineni2002bleu,nie2021coregen}.
Thus, this field of work is not helpful for us, but it could strongly benefit from an automatic assessment of CM quality:
on \cite{nie2021coregen}'s ground truth of 2,521 CMs (obviously poorly-written already filtered out),
our framework assessed 85\% to have low quality, 48\% due to missing context.

\section{Approach}
\label{sec:approach}

We present a framework that performs the first full-fledged assessment of CM quality,
including semantics and context, following
Chris Beams' guidelines (see~\cite{Beams} and Table~\ref{table:chrisBeamsRules}). 

\subsection{The Classification Pipeline}
\begin{figure*}[tb]
    \centering
        \includegraphics[width=0.65\textwidth]{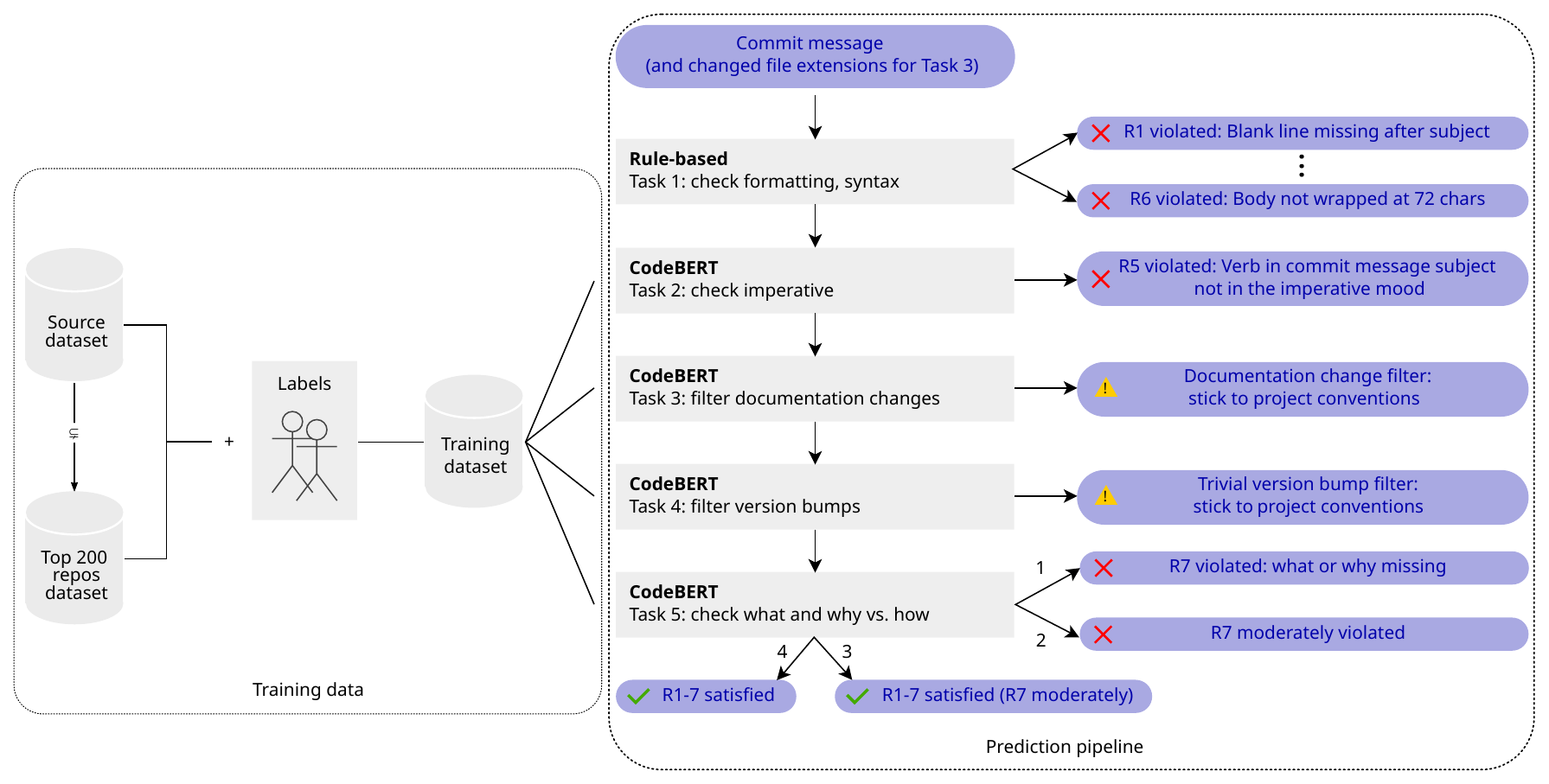}
        \caption{\label{fig:method}Training data and prediction pipeline for our novel framework for automatic CM quality assessment. It reads commit messages (and a file extension list) and outputs a rule violation warning, recommendation for project-specific conventions or for R7 improvements, or that overall CM quality is good.}
\end{figure*}

As Fig. \ref{fig:method} shows, the framework has a pipeline architecture to successively perform five tasks,
all of which take a CM as input and perform a classification.
The order of the tasks is from simplest (formatting) to hardest (semantic),
thus giving feedback as fast as possible.
With many low-quality CMs already filtered out, harder tasks can focus on CMs with potentially high quality.
Classification for the last task yields a rating between 1 (worse) and 4 (best). All other tasks perform binary classifications. 
A classification checks either a rule of \cite{Beams} or whether project-specific conventions
should be followed instead of R7.
If a rule violation is detected, the pipeline outputs a warning and exits (Tasks 1, 2 and 5).
If a project-specific convention is detected, the pipeline outputs a corresponding recommendation and exits (Tasks 3 and 4).
A recommendation is also issued when the CM has minor deficits in regards to R7 (Task 5).
So the pipeline output is either a rule violation warning, a recommendation to follow project-specific conventions or improve R7, or
\verb!R1-7 satisfied! for a successful pass through all classifiers, indicating high CM quality according to \cite{Beams}.
For warnings and recommendations issued by the pipeline, developers should review Beams' guideline \cite{Beams},
follow project-specific conventions, seek mentorship or feedback from other members of the development team or review past CMs of the project.
Therefore, the output of the pipeline directly contributes to improvement of CM quality.

Overall, our framework addresses the following tasks: 

\textbf{Task 1 (check formatting rules).} This checks whether a CM fulfills rules R1 - R4 and R6 (see Table~\ref{table:chrisBeamsRules}). It is a minor task, so we do not go into detail.

\textbf{Task 2 (check the syntactic rule).} Task 2 checks the only syntactic rule, R5: the subject line uses the imperative verb mood.
It is challenging due to syntactic ambiguities (see Table~\ref{table:chrisBeamsRules})
and the unique vocabulary of CMs \cite{Sarwar2020}.

\textbf{Task 3 (detect documentation changes) \& Task 4 (detect trivial version bump changes).}
Though not part of Chris Beams' rules, these classifiers check whether the CM describes
\begin{compactitem}
\item a documentation change, i.e., a change that does not modify functional code, only documentation like manuals and code comments (Task 3)
\item a trivial version bump, i.e., a change that merely bumps the version of the code project or one of its dependencies, but has no major consequences
  nor goals beyond the migration to the new version (Task 4)
\end{compactitem}
These rules are not part Chris Beams' guideline, but related to it:
The labeling and discussions among the 5 experts revealed that in these two cases
(documentation and trivial version bumps), changes are self-explanatory and done routinely.
The CMs for these changes need no elaborate explanation and often follow project-specific conventions.
Thus, we recommend following project-specific conventions when we detect a documentation change or
a trivial version bump change.

\textbf{Task 5 (check the semantic rule).} Task 5 checks the only semantic rule, R7: the body is used to explain what and why vs. how.
Task 5 is the main focus of this work and the most challenging task as it requires comprehension of both the CM and further context to decide which context is relevant and should be part of the CM.
Thus, Task 5 interprets R7 strictly:
It checks not only that the CM text focuses on ``what and why'' instead of ``how'', but also that it supplies sufficient ``what and why'',
i.e. all context relevant to fully understand ``what and why''.
R7 originally applies to the CM body. We extend it to the CM summary as well in correspondence to \cite{LinuxConvention} and Beams' argumentation that a single line CM can be sufficient.
Since CMs are nuanced with regard to R7, this task does not classify binarily.
Instead, a 4-point scale from 1 (violated) over 2 (moderately violated) and 3 (moderately satisfied) to 4 (satisfied) is used,
which minimizes central tendency bias (details in Sec. \ref{sec:orgb37eecb}).
Being the last task in the pipeline, its classification result for R7 is output to the user,
scores 1 and 2 being warnings that R7 is violated, score 3 meaning that all rules including R7 are met,
but recommending to further improve R7, and score 4 indicating high overall commit message quality.

\subsection{Architecture of the classification pipeline}

\textbf{Task 1} is simple, requires no ML, and is implemented as a rule-based classifier. 
Thus it is easily configurable (e.g., allowing 75 characters in the subject line~\cite{LinuxConvention} instead of 50).

For \textbf{Task 2-5}, we consider several state-of-the-art machine learning methods. Our final framework uses 
a language model (LM) based on CodeBERT (see \cite{Feng2020}) that we fine-tuned to the corresponding task.
In our evaluation in Section \ref{sec:org68dea4d}, CodeBERT was for each task either best or
among the top 3 best performing classifiers and only marginally worse than the best one (see Table~\ref{table:bestModels}).
Always using CodeBERT simplifies future multi-task learning, common pretraining, and multimodal architectures (see~\cite{Gu2021}).

For each task, the CM text is passed to the LM's pretrained tokenizer to create
input embeddings $E$.
The embeddings are input to the encoder, which produces a contextual representation $T$ summarizing
the whole CM. Pretraining with documentation and code enables $T$ to capture contextual knowledge.
Using a linear layer, $T$ is projected to the predicted label for Tasks 2-4,
and for Task 5 to the interval $[0,1]$. These are finally mapped to
\verb!R7 violated!, \verb!R7 moderately violated!, \verb!R1-7! \verb!satisﬁed (R7 moderately)!, and \verb!R1-7 satisfied! by our pipeline.

\begin{observation}
  \textbf{Contribution 1:} We propose a practical and programming language-agnostic framework for CM quality evaluation based on all rules of Beams' guideline.
\end{observation}

\section{Dataset Creation}
\label{sec:orgd1c817e}

To the best of our knowledge, there is no suitable labeled data available to train and evaluate our classification Tasks 2-5.
Thus, we -- two authors and further three software developers, all with 8+ years of industry experience across various domains, programming languages and technologies -- create suitable datasets.

\subsection{Dataset Sampling}
\label{sec:datasetsampling}

As depicted in Fig.~\ref{fig:method}, we use the following data pipeline:
\begin{compactenum}
  \item two base datasets are created: a diverse \textit{source dataset} based on 1,700 most popular repositories from GitHub and \textit{top 200 repo dataset} $\subsetneq$ \textit{source dataset} with the top 200 of those 1,700 repositories according to their quality.
  \item the  datasets to train and evaluate Task 2-5 are created out of the base datasets by evenly sampling the data out of the \textit{source dataset} and the \textit{top 200 repo dataset}.
\end{compactenum}
This data pipeline helps to focus on data for labeling and training models that are accurate and robust
in various application areas.
All datasets are open-sourced\hyperlink{sp}{$^2$}.

\subsubsection{\textbf{\textit{Source dataset}}}
We apply a sampling approach similar to Sarwar et al. \cite{Sarwar2020} and Zafar et al. \cite{Zafar2019a} to ensure a wide variety of software repositories and CM styles, e.g. no restrictions, with CM quality guidelines, Conventional Commits \cite{Conventional}, or CM templates.
Specifically, we choose the 17 most popular languages on GitHub according to the number of repositories using them. For each of the languages, we choose the most popular 100 repositories according to their number of stars and from each of those 1,700 repositories, we collect their 100 latest commits as those are contained in the data that the GitHub API returns\footnote{\label{footnote:samplingok}The balancing in our data pipeline reduces labeling effort for reaching a training set variance that is required for the model to generalize sufficiently. The balancing does not hurt validation: the predictive performance (MCC) of our best model for Task 5 only deviates by 2.54\% (STD by 2.75\%) between our and the original distribution.}.

\subsubsection{\textbf{\textit{Top 200 repo dataset}}}

Our \textit{source dataset} is highly skewed towards low quality CMs (see Sec. \ref{sec:introduction}).
Our framework should help newcomers learn to write good commit messages,
as well as teams that already take pride in writing high quality commit messages to further optimize them.
To be able to train models that predict accurately for both kind of users,
we create the \textit{top 200 repo dataset} which only contains the top 200
repositories, according to our own evaluations.

When creating all datasets to train and evaluate Task 2-5, we evenly sample
from the \textit{source dataset} (marked with the label \textit{random pool})
and from the \textit{top 200 repo dataset} (marked with \textit{good pool}),
avoiding any duplicates$^{\ref{footnote:samplingok}}$.

For Tasks 2-4, we end up with datasets of over 1,250 samples each, covering 644, 564 and 464 repositories respectively. For Task 5, we end up with a dataset of 808 samples, covering 427 repositories.

\subsection{Dataset Labeling}
\label{sec:orgb37eecb}

\subsubsection{\textbf{Task 2-4}}
Our labeled datasets for Task 2-4, annotated according to our annotation guide\hyperlink{sp}{$^2$},
contain over 1,250 samples with dichotomous annotations each, labeled by 2 experts.

\subsubsection{\textbf{Task 5}}
Since CMs are nuanced with regard to R7, we label on a 4-point scale from 1 (violated) over 2 (moderately violated) and 3 (moderately satisfied) to 4 (satisfied).
Compared to the 5-point Likert scale~\cite{likert32}, this encourages decisive evaluations and minimizes central tendency bias.
To handle uncertain cases, experts could mark commits they were uncertain about,
ensuring that we included a labeled commit message only if the corresponding expert was confident.
This approach combines the benefits of a forced-choice scale with the ability to handle uncertain cases,
resulting in a more reliable and informative labeled dataset.
Annotating a dataset for Task 5 is very challenging due to the required implicit knowledge in software development
and ambiguities of the context.
Thus, we applied the methodology depicted in Fig. \ref{fig:datasetCreationTask5}:

\begin{figure*}[tb]
\centering
\includegraphics[width=0.78\linewidth]{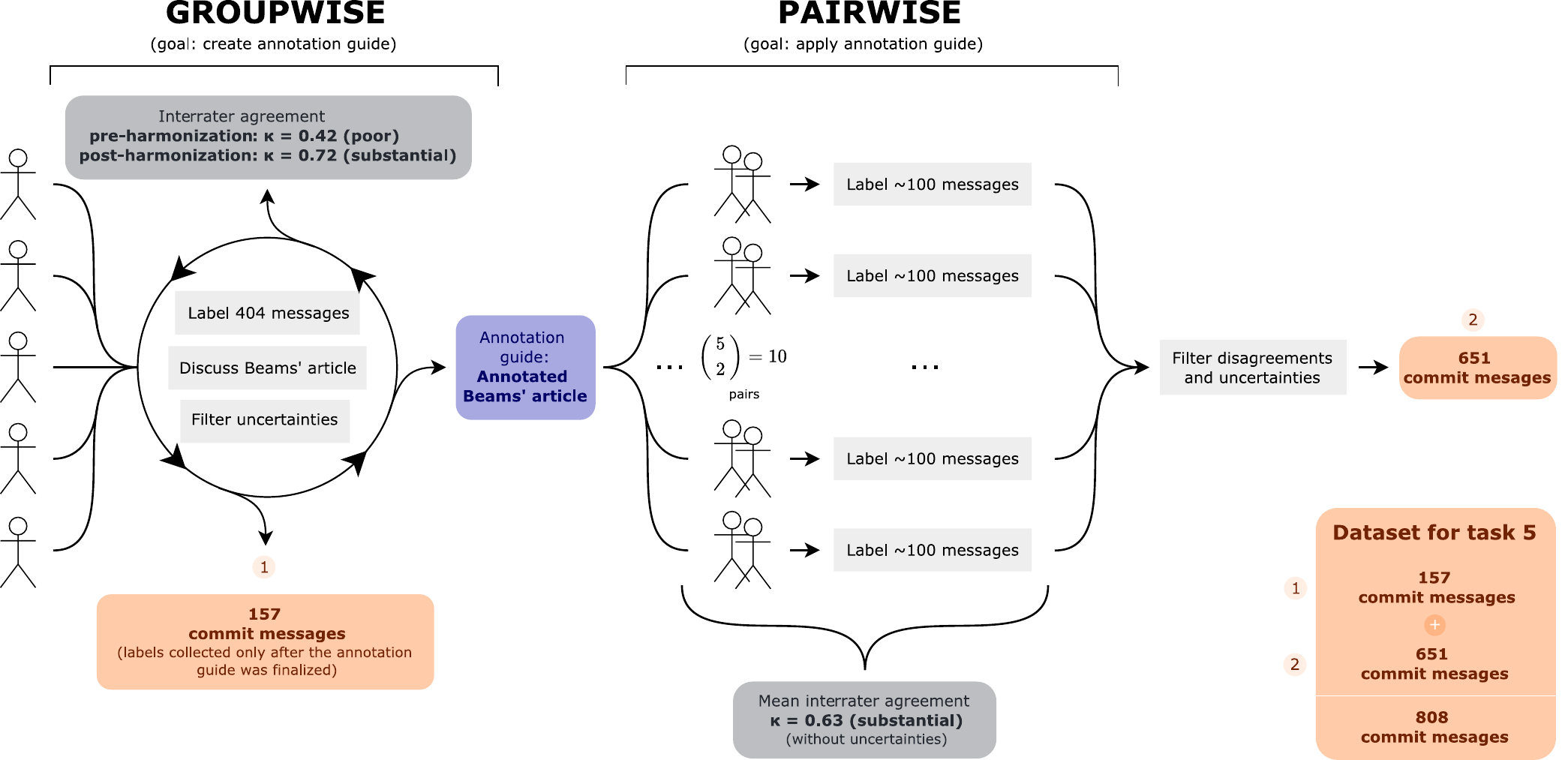}
\caption{\label{fig:datasetCreationTask5}Overview of the labeling process for Task 5.}
\end{figure*}

\textbf{Groupwise.} To ensure that our experts are all properly trained and agree on the task, we firstly conducted a feasibility study in form of a groupwise interrater agreement:
all 5 experts rated the same set of commits and conducted thorough discussions.
We estimated the amount of commits necessary for statistical significance by using the one-sample proportion in the Z-interval.
We used \(p=0.5\), a confidence interval of \(95\%\) and an error margin of \(0.05\), resulting in a minimum of \(385\) samples -- we end up using 404 commits.
Through the discussions, our agreement rose from \(\kappa=0.42\) to \(\kappa=0.72\),
i.e. from ``poor'' to ``substantial'' according to Emam et al. \cite{Emam1998}.
As a result, we created our final annotation guide\hyperlink{sp}{$^2$} in form of Chris Beams' article \cite{Beams} extended by examples and comments.

\textbf{Pairwise.} Based on the final annotation guide\hyperlink{sp}{$^2$}, experts continued to label a dataset without access to others' ratings.
To ensure the highest possible training dataset quality within what is economically feasible,
we annotated further data by splitting commits equally among $\binom{5}{2} = 10$ expert pairs.
Discussions to resolve disagreements were no longer conducted.

\textbf{Final dataset.}
For our final dataset, we filtered out commits with disagreements.
Experts marked commits they were uncertain about, which were 18\% of the commits they rated, illustrating the difficulty of this task.
This leads to 157 commits from the final phase of the groupwise interrater agreement, which have been labeled according to our final
annotation guide\hyperlink{sp}{$^2$}, and
651 commits from the pairwise interrater agreement. In total, the final dataset for Task 5 consists of 808 CMs,
407 stem from the \textit{good pool} and 401 from the \textit{random pool}.
For the pairwise interrater agreement, we have an average \(\kappa=0.63\) (substantial) for the \textit{good pool},
\(\kappa=0.60\) (moderate) for the \textit{random pool}, and \(\kappa=0.63\) (substantial) overall.
Thus, Task 5 is a feasible task and our final dataset is useful for training and evaluation.

\begin{observation}
  \textbf{Contribution 2:}
  We create the first labeled datasets for Beams' rules, including the first one to fully capture the semantics of R7.
  Representing over 400 repositories each, they are diverse.
  We ensure high-quality labels by involving 5 industry experts.
\end{observation}

\textbf{Disagreement discussion for Task 5.}
\label{sec:discussions}
During annotations, we continuously monitored \(\kappa\) agreement between experts.
We conducted 3 agreement alignment meetings in regular intervals where patterns of disagreements
were discussed and resolved.
These disagreements were mainly caused by implicit context required for some CMs:
Experts found some formulations of CMs to be ambiguous. Especially shorter messages, which naturally might not provide sufficient context, were more often subject to disagreements than longer ones. CMs that were disagreed upon were 171 characters long on average, while CMs that were agreed upon were 48\% longer with 253 characters on average. This is probably because the task is heavily context dependent and shorter CMs are more likely to contain implicit context, which can be disagreed upon.

Besides implicit context, another source for disagreements are multiple independent changes in one commit:
Some commits are not atomic, i.e., are not following the single responsibility principle (SRP)~\cite{WhoT,Chahal2018}.
This makes it harder to assess whether sufficient reason is given for the change.
The experts settle to rate such a CM positively only when the Boy Scout Rule~\cite{97things} is applied
or the message provides reasoning for all of the included logical changes.

\section{Evaluation}
\label{sec:org68dea4d}

In this section, we answer our research question
\textbf{RQ: How well can the CM quality, including semantics and context, be measured with machine learning methods?}.
For this, the model performance of various machine learning models 
used for each task is evaluated via an extensive offline evaluation.

\subsection{Evaluation setting}
\label{sec:orgf122319}

Table~\ref{table:modelsEmbeddings} lists the models we trained and evaluated, consisting of Baseline and of Transformer-based models. 

\textbf{Baselines.}
For comparison, the following baselines are used: (1) and (2)\footnote{These are the only 
  tools we found that detect the mood of the verb of a CM. They both utilize the POS tagger provided by NLTK \cite{NTLK}.},
which are tools specifically for Task 2, and (3) to (7), which have proven to perform well for a variety of NLP tasks, for Task 2-5.

\begin{compactenum}
    \item[(1)]{GitCommitBear (GCB)} \cite{GitCommitBear}, 
    \item[(2)]{Bad Commit Message Blocker (BCMB)} \cite{BadCommitMessageBlocker}
    \item[(3)]{Support Vector Machines (SVM)} trained with TF-IDF embeddings
    \item[(4)]{Random Forests (RF)} trained with TF-IDF embeddings
    \item[(5)]{FastText}, a word embedding and classification model that was on par with the performance of deep learning methods as of 2017 \cite{Joulin2017}
    \item[(6)]{Feed-Forward Neural Network (NN)} based on self-trained dense vector word embeddings
    \item[(7)]{Convolutional Neural Network (CNN)} with max pooling over time, similar to the architecture proposed by Collobert et al. \cite{RonanCollobert2017}, also based on self-trained dense vector word embeddings.
\end{compactenum}

\textbf{Transformer-based models.}
We evaluate models based on BERT \cite{Devlin2019} that utilize the encoder part of the Transformer architecture \cite{Vaswani2017}. BERT-based models have been pretrained on vast amounts of natural language text. Pretraining facilitates transfer-learning, allowing models to adapt to specific NLP tasks. BERT-based models have been preferred for classification of commit messages in research \cite{Tian22,Ghadhab2021,Sarwar2020,Feng2020}.

We evaluate the original BERT \cite{Devlin2019} itself, which is pretrained on English Wikipedia and freely available books.

Along BERT, we evaluate 2 models that enhance the architecture, parameters and training data of BERT: RoBERTa \cite{Liu2019a}, a robustly optimized version of BERT, and DeBERTa \cite{He2020}, featuring disentangled attention and an enhanced mask decoder, both trained with more data than BERT -\ DeBERTa also utilizing stories and data from Reddit, and RoBERTa news articles in addition to that.

We also evaluate DistilBERT \cite{Sanh2019}, a distilled, performant version of BERT that reduces its size by 40\% and improves its runtime by 60\% by only sacrificing 3\% of its language understanding capabilities, rendering it a compelling option for tools that are run locally on the developer's machine.

Additionally, we investigate the performance of models that have been pretrained with data coming closer to commit messages: CodeBERT \cite{Feng2020}, pretrained unimodally on programming code and bimodally on programming code and programming code descriptions, and SciBERT \cite{Beltagy2019SciBERT}, pretrained on biomedical and computer science papers.

Since our experiments with modified architectures (e.g., using Bi-LSTM) did not yield better performance, we focus on multimodality and fine-tuning various BERT-based models.

\begin{table}[tb]
\caption{\label{table:modelsEmbeddings}Classifiers and text representations of used methods.}
\centering
\begin{small}
\begin{tabular}{lll}
\toprule
Shorthand & Classifier & Text representation\\
\midrule
SVM & SVM & TF-IDF\\
RF & Random Forests & TF-IDF\\
FastText & Linear classifier & Word vectors\\
NN & Feed-forward NN & Word vectors\\
CNN & Feed-forward NN & Word vectors\\
*BERT* & Feed-forward NN & Transformer \cite{Vaswani2017} encoder \\
\bottomrule
\end{tabular}
\end{small}
\end{table}

Table \ref{table:hyperparams} lists for each model the hyperparameters that we optimized, via 10-fold random search cross-validation.
When fine-tuning BERT-based models, we vary the epochs in the range of \({5, 10, 15}\) and the learning rate in the range of \mbox{3e-5}, \mbox{4e-5}, \mbox{5e-5}. After many preliminary experiments with more hyperparameters and broader spectra, we honed in on these ranges, and on the batch-size 8.

\begin{table}[htb]
\caption{\label{table:hyperparams}Optimized hyperparameters for each model}
\begin{center}
\begin{small}
\begin{tabular}{ll}
\toprule
Method & Hyperparameters\\
\midrule
SVM & kernel, degree, C, gamma\\
RF & number of estimators, max features, max depth,\\
 & min. samples for internal node split,\\
 & min. samples required to be at a leaf node\\
FastText & parameters optimized by \cite{fastTextAutotune}\\
NN & output dimension of embedding layer,\\
 & number of hidden layers, number of units per \\
 & hidden layer, L2 regularization parameter \(\lambda\)\\
CNN & kernel size, output dimension of embedding layer,\\
 & number of hidden layers, number of units per \\
 & hidden layer, L2 regularization parameter \(\lambda\)\\
*BERT* & learning rate, epochs\\
\bottomrule
\end{tabular}
\end{small}
\end{center}
\end{table}

\subsection{Evaluation results}

Table \ref{table:bestModels} summarizes the best 3 BERT-based and the best 3 baseline models for each of our tasks.
Besides precision, recall, and F$_1$, the Table~\ref{table:bestModels} lists the MCC score (Matthews' correlation coefficient)~\cite{Matthews1975} and its standard deviation. 
All baseline methods were significantly outperformed by methods based on the BERT architecture.
For Task 2, the difference was largest, for Task 4 smallest.
The performance of the best three BERT architectures differed by at most 5\% for each task.
Since CodeBERT was the best
(except for Task 4 where DeBERTa was only 1\% better),
we decided to use CodeBERT as model for all Tasks 2-5, since that simplifies the pipeline and makes it more flexible for future improvements (multimodality, multi-task learning, own pretraining).
CodeBERT's higher performance compared to other BERT-based models is likely due to its smaller covariate shift between its pretraining data and our data.

\begin{table}[tb]
\caption{\label{table:bestModels}Performance of the respective 3 best unimodal BERT-based and respective 3 best simple baseline models for each task (Task 1 is rule-based and thus uses no model that needs to be evaluated), and BCMB \cite{BadCommitMessageBlocker} and GCB \cite{GitCommitBear} for Task 2.}
\centering
\begin{tabular}{rlrrrrr}
\toprule
Task & Method & MCC & F$_1$ & Precision & Recall & MCC STD\\
\midrule
2 & GCB & 0.478 & 0.651 & 1.000 & 0.373 & 0.046\\
 & BCMB & 0.563 & 0.745 & 0.938 & 0.551 & 0.036\\
\cmidrule(r){2-7}
 & RF & 0.622 & 0.810 & 0.808 & 0.815 & 0.077\\
 & fastText & 0.705 & 0.850 & 0.890 & 0.802 & 0.070\\
 & NN & 0.750 & 0.873 & 0.889 & 0.856 & 0.053\\
\cmidrule(r){2-7}
 & SciBERT & 0.949 & 0.973 & \textbf{0.980} & 0.967 & 0.031\\
 & BERT & \textbf{0.958} & \textbf{0.978} & \textbf{0.980} & 0.977 & 0.027\\
 & CodeBERT & \textbf{0.958} & \textbf{0.978} & 0.978 & \textbf{0.978} & \textbf{0.022}\\
\midrule
3 & fastText & 0.709 & 0.853 & 0.876 & 0.809 & 0.082\\
 & RF & 0.718 & 0.857 & 0.887 & 0.811 & 0.062\\
 & SVM & 0.723 & 0.857 & 0.910 & 0.788 & 0.059\\
\cmidrule(r){2-7}
 & RoBERTa & 0.818 & 0.906 & 0.903 & 0.911 & 0.053\\
 & DeBERTa & 0.823 & 0.909 & 0.905 & 0.914 & 0.048\\
 & CodeBERT & \textbf{0.840} & \textbf{0.918} & \textbf{0.913} & \textbf{0.924} & \textbf{0.042}\\
\midrule
4 & RF & 0.822 & 0.913 & 0.913 & 0.912 & 0.448\\
 & fastText & 0.847 & 0.922 & 0.955 & 0.890 & 0.051\\
 & NN & 0.869 & 0.934 & 0.934 & 0.938 & 0.041\\
\cmidrule(r){2-7}
 & CodeBERT & 0.897 & 0.946 & \textbf{0.956} & 0.938 & \textbf{0.022}\\
 & SciBERT & 0.898 & 0.947 & 0.954 & 0.941 & 0.025\\
 & DeBERTa & \textbf{0.906} & \textbf{0.952} & 0.954 & \textbf{0.949} & 0.038\\
\midrule
5 & CNN & 0.529 & 0.724 & 0.836 & 0.720 & 0.123\\
 & NN & 0.532 & 0.734 & 0.795 & 0.783 & 0.111\\
 & FastText & 0.567 & 0.762 & 0.823 & \textbf{0.813} & 0.083\\
\cmidrule(r){2-7}
 & DistilBERT & 0.676 & 0.814 & 0.823 & 0.809 & \textbf{0.069}\\
 & SciBERT & 0.686 & 0.818 & 0.832 & 0.809 & 0.070\\
 & CodeBERT & \textbf{0.703} & \textbf{0.829} & \textbf{0.841} & \textbf{0.820} & 0.070\\
\bottomrule
\end{tabular}
\end{table}

\textbf{Task 2.} We used the additional baseline methods GitCommitBear \cite{GitCommitBear} and Bad Commit Message Blocker \cite{BadCommitMessageBlocker}, which are specialized for this task and therefore listed in Table~\ref{table:bestModels} in spite of performing poorly. Both are outperformed by all other baseline methods.
Their low performance is likely due to the simple n-gram model \cite{NLTKPerceptronTagger} used to train the POS tagger,
which is unable to capture advanced context, and due to the covariate shift between their training and our evaluation data (see \cite{NLTKPosIssues}). 

\textbf{Task 3 and 4.} Task 4 is the only task where CodeBERT is not the best model, but it is only slightly worse (0.006 F$_1$) than the best. Table~\ref{table:bestModels} contains our unimodal CodeBERT model that is trained with the CM text alone. 
In Table~\ref{table:docsMultimodal}, we list the performance of our bimodal CodeBERT model for Task 3, which is trained with the CM text and prepended counts of the extensions of changed files (e.g., $md3 java1$ for 3 changed Markdown files and 1 changed Java file).
Compared to our unimodal CodeBERT model for Task 3, bimodality improves the performance of the model significantly, from F$_1=0.918$ to F$_1=0.954$. %
Thus, extension counts are a viable additional feature for classification of documentation changes. Our final pipeline incorporates bimodality in Task 3.

\begin{table}[tb]
\caption{\label{table:docsMultimodal}Unimodal (CM) and bimodal (+extension counts) performance of CodeBERT for Task 3}
\centering
\begin{tabular}{rrrrrr}
\toprule
Modality & MCC & F$_1$ & Precision & Recall & MCC STD\\
\midrule
unimodal & 0.840 & 0.918 & 0.913 & 0.924 & 0.042\\
+ extension counts & \textbf{0.911} & \textbf{0.954} & \textbf{0.946} & \textbf{0.963} & \textbf{0.040}\\
\bottomrule
\end{tabular}
\end{table}

We experimented with other categorical and numerical features as part of the textual BERT model input, like file extensions, change lengths and an estimation of the change diff entropy, but only the introduction of extension counts to the model input of Task 3 lead to a significant improvement in the classification performance.
Overall, for Task 4, multimodality did not improve performance.

\textbf{Task 5.}
Compared to other tasks, CodeBERT won most clearly for Task 5 with an F$_1$ difference of 0.011 compared to the second best, SciBERT. 
As for Task 4, multimodality did not improve performance for Task 5.

\begin{figure}[tbhp]
\centering
\includegraphics[width=0.78\linewidth]{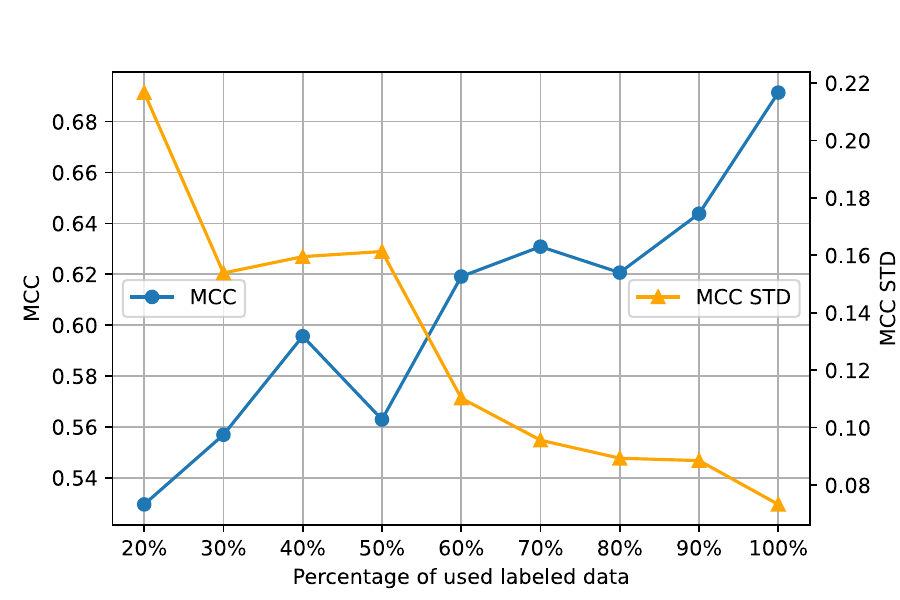}
\caption{\label{fig:dataSizeEffect}Effect of data size on the predictive performance of our best model for Task 5 where 100\% corresponds to the 808 commits in our sample evaluated by a 10-fold CV.}
\end{figure}

Being the most challenging task, we analyze the performance in more detail:
Firstly Table~\ref{table:bestModelMetrics} summarizes F$_1$, precision, and recall for both positive (minority class, Task 5 ratings $< 2.5$) and negative (majority class, Task 5 ratings $\geq 2.5$) class for our best model, CodeBERT.
Secondly, Table \ref{table:tianEtAlComparison} compares the performance of the best model (BERT) from  Tian et al. \cite{Tian22} and our best model (CodeBERT).
We evaluate both models on our own dataset, due to the deficits of Tian et al.'s dataset, see Sec.~\ref{sec:relatedwork}. Our model is evaluated using 10-fold cross-validation,
while their model was trained on their own dataset and then evaluated on our dataset.
Comparing both models on our dataset, our model has a 35\% better F$_1$ score, 32\% better precision, and 53\% better recall.
Evaluating their model on our dataset instead of their dataset, their (see \cite{Tian22}) positive class F$_1$ score drops from 77.6\% to 73.6\% (5\%),
their negative class F$_1$ score from 73.9\% to 61.5\% (17\%).
This shows that their model does not perform well for other programming languages than Java, domains outside high quality open-source web communication,
a semantically meaningful interpretation of rule R7 (see Sec.~\ref{sec:relatedwork}), and is thus unsuitable for general use.
In contrast, our approach does not have these limitations, as the F$_1$ score of 82.9\% shows.
The performance can likely be improved further by increasing Task 5's dataset (see Sec.~\ref{sec:futurework}),
as the plot of the predictive performance, depending on the dataset size, shows in Fig. \ref{fig:dataSizeEffect}.

\begin{table}[tbhp]
\caption{\label{table:bestModelMetrics}Performance of CodeBERT for Task 5.}
\centering
\begin{tabular}{@{}llll@{}}
\toprule
Class    & F$_1$    & Precision & Recall \\ \midrule
Negative & 0.872 & 0.866     & 0.880  \\
Positive & 0.829 & 0.841     & 0.820  \\ \bottomrule
\end{tabular}
\end{table}

\begin{table}[]
\caption{\label{table:tianEtAlComparison} Performance comparison with Tian et al.'s pretrained classifier \cite{Tian22} on our dataset}
\centering
\begin{tabular}{@{}lllll@{}}
\toprule
Method                             & F$_1$    & Precision & Recall & Accuracy \\ \midrule
Tian et al. (BERT) & 0.615 & 0.658     & 0.577  & 0.759 \\
We (CodeBERT) & \textbf{0.829} & \textbf{0.841}     & \textbf{0.820} & \textbf{0.854}  \\ \bottomrule
\end{tabular}
\end{table}

\begin{figure}[tbhp]
    \centering
        \includegraphics[width=0.44\textwidth]{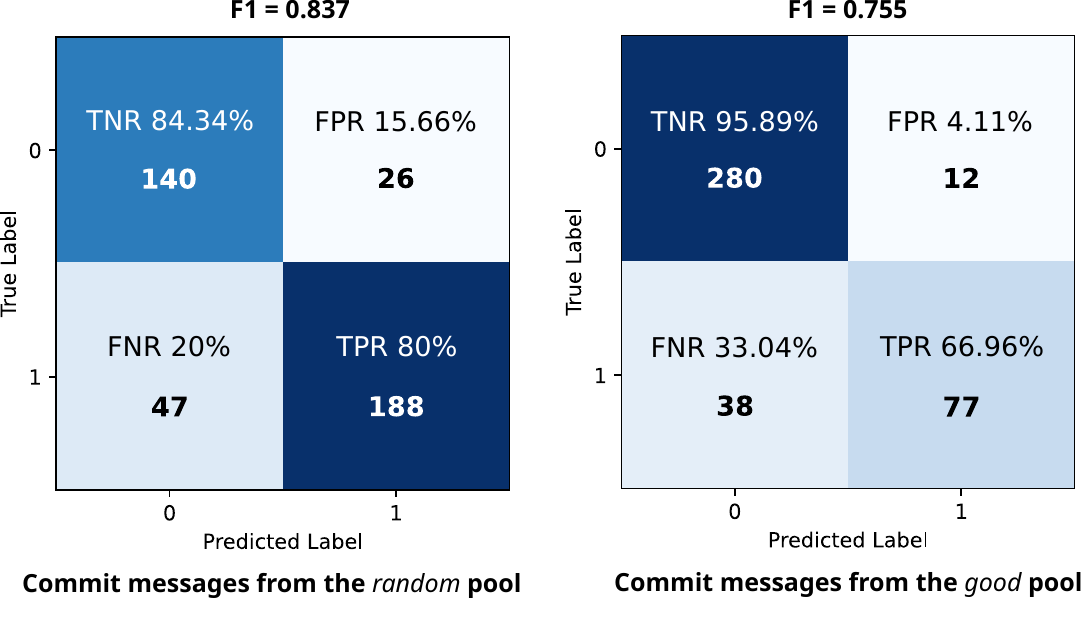}
        \caption{\label{fig:rand-good-cm}Performance of our CodeBERT trained and evaluated with 10-fold CV, with evaluation results separated for commits from the \textit{good} and \textit{random} pools.}
\end{figure}

\textbf{Task 5: error analysis.}
\label{sec:org5838c29}
Since Task 5 is the most challenging and yields the lowest performance among our tasks, we conduct an error analysis:
Firstly, we see in Table~\ref{table:bestModelMetrics} that our model is weakest on the positive class recall with 0.82.
With a FNR of 0.18, our model classifies rather lax and predicts 18\% of bad commit messages as good.
But for a quality assurance tool, being too lax is better than being too strict since a too high FPR leads to
noise, manual work, distrust, and eventually rejection of the tool.
Nonetheless we look into the cause for our lax model:
For Fig.~\ref{fig:rand-good-cm}, we conduct another 10-fold cross-validation to create confusion matrices on the \textit{random pool} resp. the \textit{good pool}
(see Sec.~\ref{sec:datasetsampling}). As expected, the FNRs are higher than the FPRs.
The model has a significantly higher TPR in the \textit{random pool} (80\%) than in the \textit{good pool} (67\%),
likely due to many more obviously bad CMs in the \textit{random pool}.
As consequence from the high FNR in the \textit{good pool}, the overall positive class recall is relatively low.
So the overall positive class recall would be better with our data pipeline sampling the training set only from the \textit{random pool}
instead of evenly from the \textit{random pool} and \textit{good pool}.
We believe even sampling was the right choice because our model should also be helpful
for teams that already take pride in writing high quality commit messages. 
With a training set from the \textit{random pool} only, the model would have a harder time to generalize to the \textit{good pool},
and Fig.~\ref{fig:rand-good-cm} shows that the model with even sampling already performs worse on the \textit{good pool} than on the \textit{random pool}.
All in all, we conclude from Fig.~\ref{fig:rand-good-cm} that the model performance for the \textit{good pool}, though reasonable, could be improved by
labeling and sampling more CMs from the \textit{good pool}.

\textbf{Runtime performance.}
\label{sec:org849769e}
The CodeBERT models are about 500 MB in size and evaluating a sentence takes about 7 seconds on a modern i7-1068NG7 CPU without utilizing GPU support.
This may not be sufficient for a satisfactory user experience for real-time evaluation while the CM is being typed.
When exporting the model to the ONNX \cite{onnx} format and thus profiting from hardware optimizations, the runtime of a single model prediction in the pipeline on the CPU can be reduced significantly to 0.02 seconds. Thus, we use ONNX.

In conclusion, given the high predictive performance of F$_1 =$ 82.9\% for our CodeBERT model for Task 5,
our answer to \textbf{RQ: How well can the CM quality, including semantics and context, be measured with ML methods?} is: sufficiently high for practical use.

\begin{figure}[tbhp]
  \centering
  \includegraphics[width=0.45\textwidth]{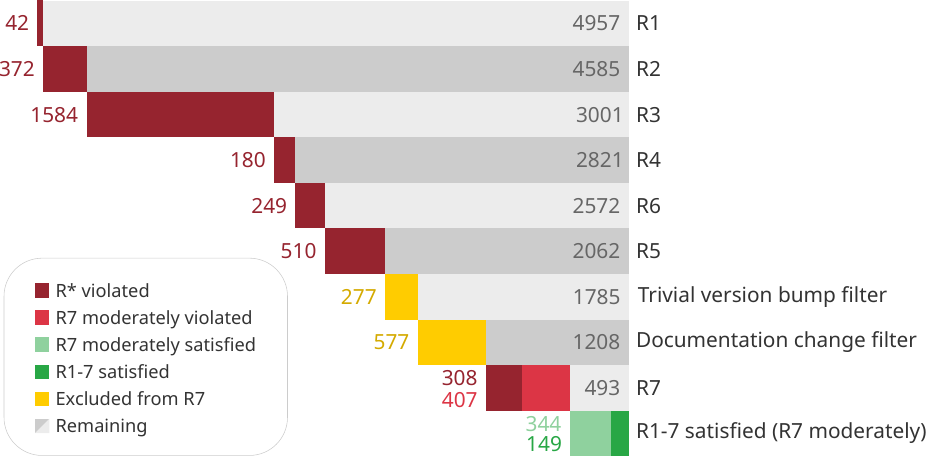}
  \caption{\label{fig:application5k}Our pipeline classifying 5,000 randomly sampled commits}
\end{figure}

For illustrative purposes, we demonstrate the application of our pipeline to a random sample of 5,000 CMs in Figure \ref{fig:application5k}.

\begin{observation}
  \textbf{Contribution 3:} Using our datasets, we fine-tune current NLP methods. They achieve state-of-the-art and can measure CM quality -- including semantics and context -- sufficiently well for practical use, answering our research question positively.
\end{observation}

\section{Discussion}
\label{sec:discussion}

\subsection{Predictors of CM quality}

We use our tool to assess various repositories (repos for short) and look into multiple aspects of software engineering
to investigate predictors of CM quality, with linear regression and Pearson correlations
(since outliers occurred rarely and in those cases the Spearman correlation coefficient was only insignificantly different). All reported correlation coefficients have $p<0.05$.

\textbf{Commit-based predictors.}
Firstly, we scrape a large \textit{Commit Guru dataset}\hyperlink{sp}{$^2$} from \cite{cg_hp}. 
It comprises 482 repos, each with its full commit history, totaling 4,103,489 commits,
and each with many per-commit metrics. Commit Guru~\cite{rosen15} derives
these metrics as estimates for aspects like
the number of developers involved, their experience,
whether the commit introduces a bug (\textit{bug-inducing}), and various aspects regarding time.
Since there are 16 numerical metrics, we firstly analyze their relationship with R7 ratings
with an OLS multiple linear regression: it is significant and the per-commit metrics do correlate with R7 ratings
(\textit{F-statistic}$\,=17530$ with $p=0$, each predictor's $p<0.05$),
but with very high variance and each individual per-commit metric only has a very weak correlation with R7 ratings
($R^2=0.068$, prediction intervals cover the full R7 rating scale).
We perform another OLS multiple linear regression on per-repo averages across all 482 repos,
to check whether these per-repo metrics, e.g. the average number of developers involved in a repo,
have a stronger effect on the averaged R7 ratings.
This regression is also significant, and at least 6 per-repo metrics correlate with R7 ratings
(\textit{F-statistic}$\,=10.79$ with $p=0$, and $p<0.05$ for 6 predictors),
but still with relatively high variance and each individual per-repo metric has no or a very weak correlation with R7 ratings
($R^2=0.258$,  prediction intervals cover half of the R7 rating scale).
These weak correlations might be due to software engineering being complex,
with many interacting explanatory variables --
regressions have very high condition numbers of 
\mbox{1.01e+16} (per-commit) resp. \mbox{2.25e+05} (per-repo).
Consequently, we will only look at a few specific relationships in the \textit{Commit Guru dataset}.
\kuerzungskandidat{A more elaborate analysis on these predictors is future work.}

\textbf{Correlations with commit-based predictors.}
Firstly, we investigate the per-commit relationship between R7 ratings and \textit{bug-inducingness},
which is estimated with an SZZ-like algorithm~\cite{zeller05} that detects bug-fixing commits via keyword detection in CMs,
from which bug-inducing commits are derived via git blame \cite{rosen15}.
As expected after our regression analysis, we only get a weak correlation,
surprisingly a positive one.
Computing the per-repo relationship instead, i.e. the correlation between the percent of \textit{bug-inducing} commits
and average R7 ratings leads to a similar result,
as does the relationship between commit's \textit{bug-inducingness}
and their average R7 ratings so far.
These are likely due to confounding variables:
writing better commit messages leads to better keyword detection and
thus better detection of buggy commits;
and more complex environments cause more bugs, but also demand better CMs (see moderate correlation $r_\text{systemsPL}$ below).

Thus we consider other, per-repo predictors. Firstly, we compute the correlation between the per-repo average R7 rating
and the lifetime of the repo, measured by the time span between the repo's first and last
commit\footnote{The average over all repos is 8.2 years, its standard deviation 5.2 years.}:
we get a moderate positive correlation $r_\text{lifetime}=0.30$.
If $r_\text{lifetime}$ is due to a causal relationship, it is not from lifetime to R7 ratings, since the R7 ratings are roughly constant over time, but from R7 ratings causing more maintainable repos and thus higher sustainability, i.e. higher lifetime. 
Secondly, we consider team size as predictor and compute a weak positive correlation $r_\text{team}=0.28$
between the average R7 ratings and the average number of developers,
likely due to the increased necessity of communication in larger teams.

\textbf{Correlation with more general predictors.}
Having at most moderate correlations on commit-based metrics,
we consider more general predictors: programming language, developers, companies.
We use our \textit{source dataset} (see Sec.~\ref{sec:datasetsampling}),
which comprises 1,700 repos with 100 commits each,
covering 17 programming languages (PLs) and many companies.

A repository has the main PL $P$ if more than 50\% of the files changed by its commits are in $P$.
Table~\ref{table:programminglanguages} lists the main PL,
the average R7 rating, the age of the PL,
and the average age of the developers of the PL~\cite{PLbyAge}. 
This grouping leads to the following correlations:
\begin{compactitem}
\item projects using younger PLs have lower average R7 rating,
  with a strong correlation $r_\text{agePL}=0.83$
\item projects in a systems PL (C, C++, Go) have a higher average R7 rating,
  with a moderate correlation $r_\text{systemsPL}=0.58$,
  likely because developers know that low-level programs %
  interface with a complex environment outside of the software system and thus 
  demand thorough descriptions of that complex context in the CMs.
\end{compactitem}

\begin{table}[tb]
\caption{\label{table:programminglanguages}PLs with average R7 score, age, average age of developers}
\centering
\begin{small}
\begin{center}
\begin{tabular}{lllll}
\toprule
\textbf{Main PL} & \textbf{Score} & \textbf{Age of PL}  & \textbf{Age of Dev}\\
\midrule
C &	3.34 & 51 & 31.7\\
C++ &	3.14 &	38 &	32.0\\
Perl &	2.89 &	35 &	40.3\\
Ruby &	2.80 &	28  &	34.2\\
PHP &	2.77 &	28  &	32.5\\
Objective-C &	2.76 &	38 &	34.6\\
Python &	2.75 &	32 &	30.7\\
JavaScript (JS) &	2.68 &	27 &	32.5\\
Java &	2.60 &	28 &	32.3\\
Go &	2.58 &	14 &	32.1\\
Scala &	2.51 &	18 &	33.3\\
Lua &	2.43 &	29 &	33.7\\
TypeScript (TS) &	2.37 &	10 &	30.6\\
Swift &	2.34 &	9 &	30.8\\
C\# &	2.15 &	23 &	33.8\\
CoffeeScript &	2.12 & 13 & -\\
\bottomrule
\end{tabular}
\end{center}
\end{small}
\end{table}

To investigate whether the strong correlation $r_\text{agePL}$ is caused by developer demographics,
we estimate the experience of a developer $d$ at the time they created commit $c$
by getting (a) the time span between $d$'s first GitHub commit and $c$,
and (b) the number of GitHub commits of $d$ up to $c$.
On our dataset, there is only a weak positive correlation $r_\text{experienceDev} < 0.08$ between R7 ratings and (a) resp. (b)\footnote{This is in line with the weak correlation $r=0.09$ between R7 ratings and the per-repo developer experience metric of the \textit{Commit Guru dataset}.}.
Additionally, the average developer age estimated for each PL $P$ (see Table \ref{table:programminglanguages}) has
a weak correlation $r_\text{ageDev}=0.13$ with $P$'s average R7 rating.

Is the strong correlation $r_\text{agePL}$ caused by team culture?
Lacking a metric for company culture, we group the commits into their company workplace, i.e. owner, of the corresponding repository:
The average R7 rating for each owner is listed in Table~\ref{table:companies} 
(Big-4 is Amazon, Apple, Meta, Alphabet (Goog); Big-5 is Big-4 and Microsoft (MS);
Big-tech is Big-5 and Netflix (Nflx), Snap, Twitter (Twtr), and Uber; everything else is Non-big-tech).
This shows a strong relationship between company culture and CM quality.

\begin{table}[tb]
\caption{\label{table:companies}(Groups of) corporations with their CM quality}
\begin{small}
\begin{flushleft}
\begin{tabular}{l||l|l|l|l|l|l|l}
\toprule
\textbf{Corp} & \textbf{Uber} & \textbf{Apple}  & \textbf{Twtr} & \textbf{Meta} & \textbf{Goog} & \textbf{Nflx} & \textbf{MS}\\
\midrule
\textbf{Score} & 3.92 & 3.56 & 3.43 & 3.38 & 3.20 & 2.70 & 2.68\\
\bottomrule
\end{tabular}

\vspace{1em}

\begin{tabular}{l||l|l|l|l}
\toprule
\textbf{Group} & \textbf{Big-4} & \textbf{Big-5}  & \textbf{Big-tech} & \textbf{Non-big-tech}\\
\midrule
\textbf{Score} & 3.37 & 3.19 & 3.24 & 2.67\\
\bottomrule
\end{tabular}
\end{flushleft}
\end{small}
\end{table}

In summary, a CM quality assessment with our tool is not only useful for improving CMs,
but can also reveal other software engineering aspects, e.g. about the developer culture.
The per-repo correlations with average R7 ratings are
\begin{compactitem}
\item for PLs: a strong correlation of $r_\text{agePL}=0.83$ with the main PL of the repo,
  and a moderate correlation $r_\text{systemsPL}=0.58$ with the system level of the PL,
\item for developers: weak correlations of $r_\text{ageDev}=0.13$ with the estimated average age of the developers
  and $r_\text{experienceDev}\leq 0.09$ with the estimated experience of the developers,
  and a moderate correlation of $r_\text{team}=0.28$ with the team size,
\item for sustainability: a moderate correlation of $r_\text{lifetime}=0.30$ with the lifetime of the repo.
\end{compactitem}

\subsection{Threats to validity}
\label{sec:threats}

\textbf{Threats to construct validity.}
Fig. \ref{fig:dataSizeEffect} indicates that training with more data can improve the predictive performance of our model for Task 5,
which we already started working on (see next section).
Our data collection was limited by the experts' availability and the high labeling effort for Task 5.

\textbf{Threats to internal validity.}
CodeBERT's text input is limited to 512 word(-parts)~\cite{Sennrich2016}, which can cause attrition or deformation (e.g., clipping). But this happened only for extremely long CMs, in less than $0.4\%$ of the cases.

\textbf{Threats to external validity.}
Some developers neglect writing high quality CMs, arguing that their workflow is issue tracker centric. But software maintainability is improved by having both issues and CMs of high quality. The CM history is independent of the issue tracking tool, closer to the code and developer, and accessible by many tools, e.g. IDEs.
Furthermore, there is always a one-to-one relationship between commits and CMs, but this is often not the case between commits and issues. 
Finally, many CMs in our dataset had dead PR- or issue-links, pointed to issues without any useful text or to issues with information hidden in too much text
or behind links (see e.g. Table \ref{table:chrisBeamsRules}).
Thus the git history should contain the relevant information directly.

\subsection{Future Work}
\label{sec:futurework}

\textbf{Additional labels.} We started to annotate further CMs with additional labels, covering the following
aspects: “why“ sufficient, “what“ sufficient, “how“ not distracting, CM matches code change.
Differentiating these aspects is quite complex and time consuming,
but we plan to gain further insights and performance improvements from having more data and more fine-grained labels.

\textbf{Parallel rule checking.} Our pipeline processes a CM sequentially through our classification models.
A parallel architecture would enable faster evaluation of high quality CMs, and multiple warnings at once for low quality ones.
However, it would have required experts to also label CMs for the challenging rule R7 that fail the simpler rules R1 to R6 for various reasons. But many of those CMs are hard for the experts to understand and label because they have confusing formatting (filtered out by Task 1), follow project-specific conventions (filtered out by Task 3 and 4), or text that does not clarify whether it is about the change or about the context and the way things worked before the change (filtered out by Task 2). Due to the higher variance of those CMs, the model for R7 would also need more training data to achieve a certain performance.

\textbf{User study.} We plan a user study to identify possibilities for improvement.
There is already interest from several medium and large companies to participate.

\textbf{Pretraining BERT.} CodeBERT likely performs better than other BERT-based models because it has a smaller covariate shift
between its pretraining data and our fine-tuning data.
There is some covariate shift left even for CodeBERT, so
pretraining a BERT-based method with pairs of CMs and matching code diffs
will likely further improve the performance, by achieving advanced comprehension of code diffs
and thus improve the predictive performance for our tasks.
But pretraining is expensive, as well as data- and time-consuming, and thorough experiments on pretraining tasks on commits are needed.

\section{Conclusion}
\label{sec:conclusion}

The quality of CMs is crucial for software maintenance and evolution. Thus, we considered how well CM quality can be assessed, and created a framework for automatically assessing CM quality based on several criteria, using state-of-the-art ML methods. 
Our contributions are all open-sourced\hyperlink{sp}{$^2$}.

Specifically, we developed and provided large \textbf{datasets} of labeled CMs for the detection of imperative mood (Task 2), of version bump (Task 3), and of documentation changes (Task 4). Furthermore, we designed a high-effort dataset of labeled CMs for evaluating their semantic quality based on context and meaning (Task 5 for Chris Beams Rule 7 about ``what and why vs. how''). These datasets are also useful for benchmarking and other tasks, such as research about CM generation.
We then provided thorough \textbf{evaluations} of our full-fledged framework, consisting of 4 classification tasks. To this end, we considered 7 baselines and 5 deep learning-based
models. The best performing models (BERT-based trained on our datasets) set the state-of-the-art for all our classification tasks, with the following F$_1$ scores: 97.8\% for Task 2 (formerly 74.5\%), 95.4\% for Task 3, 95.2\% for Task 4, 82.9\% for Task 5 (formerly 61.5\%).
Finally, we created our framework as a \textbf{tool} for practitioners, which can be run locally or in the cloud, and comprises a pipeline to automatically assess the CM quality on all levels (format, syntax, semantics), checking all rules of the CM guideline by Chris Beams \cite{Beams}, the most popular one.

The evaluation demonstrates that our open-source framework can automatically assess the quality of CMs, including semantics and context, sufficiently well for practical use.
Our framework can be used for assessing projects and company culture, or integrated in a software development process. For instance, as an IDE plugin for immediate validation while a CM is being written, or as commit hook~\cite{Itk2021} for a simple solution that still prevents commits with low quality messages from being merged, or in the CI as an automatic CM quality reviewer. 
There is already interest from medium and large companies to participate in our planned user study.
A CI integration in the background, e.g., of GitHub projects with GitHub Actions \cite{ghActions}, can lead to a broad adoption and thus to a shift towards more maintainable and faster evolving software.

\section*{Acknowledgment}
We thank Jochen Krause, CEO of Innoopract Informationssysteme GmbH, for his big support with knowledge and investment of 5 experts for labeling.

Furthermore, we acknowledge support by the state of Baden-Württemberg through bwHPC and by KI-Lab Lübeck for provided infrastructure.

\bibliographystyle{plain}
\bibliography{thesis}
\clearpage

\end{document}